\documentclass[preprint2,numberedappendix,iop]{emulateapj-rtx4}
\usepackage{graphicx}
\usepackage{bm}
\usepackage{natbib}
\usepackage{ulem}
\newcommand{\approptoinn}[2]{\mathrel{\vcenter{
        \offinterlineskip\halign{\hfil$##$\cr
        #1\propto\cr\noalign{\kern2pt}#1\sim\cr\noalign{\kern-2pt}}}}}

\usepackage{color}

\newcommand{\Eq}[1]{Equation~(\ref{#1})}
\newcommand{\Eqs}[2]{equations~(\ref{#1}) and~(\ref{#2})}

\newcommand{\Fig}[1]{Figure~\ref{#1}}
\newcommand{\Tab}[1]{Table~\ref{#1}}

\graphicspath{{./fig/}{./png/}}

\begin{document}
\title{Long-Term Evolution of the Sun's magnetic field during Cycles 15--19\\
based on their proxies from Kodaikanal Solar Observatory
}
\medskip
\author{Alexander  V. Mordvinov$^1$, Bidya Binay Karak$^2$, Dipankar Banerjee$^{3,4,5}$, Subhamoy Chatterjee$^6$, Elena M. Golubeva$^1$, Anna I. Khlystova$^1$}
\affiliation{$^1$Institute of Solar-Terrestrial Physics, Irkutsk, 664033, Russia}
\email{avm@mail.iszf.irk.ru, karak.phy@iitbhu.ac.in}
\affiliation{$^2$Department of Physics, Indian Institute of Technology (Banaras Hindu University), Varanasi, India}
\affiliation{$^3$Aryabhatta Research Institute of Observational Sciences, Nainital 263000, Uttarakhand}
\affiliation{$^5$Indian Institute of Astrophysics, Koramangala, Bangalore 560034, India}
\affiliation{$^6$Southwest Research Institute, 1050 Walnut St $\#$300, Boulder, CO 80302, USA}
\date{\today}

\begin{abstract}
The regular observation of the solar magnetic field is available only for 
about
last five cycles. 
Thus, to understand the origin of the variation of the solar magnetic field, it is essential to reconstruct the magnetic field for the past cycles, utilizing other datasets. Long-term uniform observations for the past 100 years as recorded at the Kodaikanal Solar Observatory (KoSO) provide such opportunity. We develop a method for the reconstruction of the solar magnetic field using the synoptic observations of the Sun's emission in the Ca~II~K and H$\alpha$ lines from KoSO for the first time. The reconstruction method is based on the facts that the Ca~II~K intensity correlates well with the unsigned magnetic flux, while the sign of the flux is derived from the corresponding H$\alpha$ map which provides the information of the dominant polarities. Based on this reconstructed magnetic map, we study the evolution of the magnetic field in Cycles 15--19. We also study bipolar magnetic regions (BMRs) and their remnant flux surges in their causal relation. Time-latitude analysis of the reconstructed magnetic flux provides an overall view of magnetic field evolution: emergent magnetic flux, its further transformations with the formation of unipolar magnetic regions (UMRs) and remnant flux surges. We identify the reversals of the polar field and critical surges of following and leading polarities.  We found that the poleward transport of opposite polarities led to multiple changes of the dominant magnetic polarities in poles. Furthermore, the remnant flux surges that occur between adjacent 11-year cycles reveal physical connections between them.
\end{abstract}
\maketitle


\section{Introduction}
\label{sec:int}
The long-term measurement of the Sun's magnetic field
\citep{Petrie15,Riley14} is essential to understand the origin of
the solar magnetic cycle as well as the causes of various processes
related to solar activity \citep{Jiang2014,Kar14a,Cha20}. Guided by the early
measurements of the solar magnetic field \citep{Babcock1955}, \citet{Babcock1961} 
and \citet{Leighton1964}
provided the basic concept of the mechanism for the generation of
the solar magnetic cycle. According to this concept, the poloidal
component of the Sun's magnetic field is generated through the decay
and dispersal of tilted bipolar magnetic regions (BMRs) at the low
latitudes. 
The poloidal magnetic field is transported towards the
higher latitudes and then downward to the deeper convection zone through meridional circulation and turbulent diffusion. 
Capturing this idea, the
Surface
Flux Transport (SFT) \citep{Wang1989,Jiang2014} and the Babcock-Leighton type dynamo models \citep{Kar14a,Cha20}
have been constructed which reproduce the magnetic field evolution on the solar surface resonably well.

Detailed comparison of the magnetic field from
these models with that of the observation is essential to constrain
the parameters of the model and thus to understand the mechanism
of the solar dynamo. 
However, regular digitalised, full-disk measurements of the Sun's magnetic
field available only in 1967 and that span for about last five cycles
(Cycles 20--24).
Since Cycle 21, the level of magnetic
activity constantly decreased. At the beginning of the last century
also, the level of solar activity was low, however then, the activity
increased, reaching the highest amplitude in Cycle 19. 
However, due to the complexity of the
dynamo mechanism and the unavailability of magnetic field measurements,
the nature of secular changes in the solar activity 
is poorly understood. Thus, it is of crucial importance to develop methods
for the reconstruction of the solar magnetic field using indirect
datasets over the past century.

The imprints of active regions (ARs) and the unipolar magnetic
regions (UMRs, which are formed from the predominantly following
polarities and are transported poleward due to meridional flow
\citep{Babcock1955,Petrie2015} are also observed in the Sun's
chromosphere. 
Long-term observations from MWO were compiled in supersynoptic charts that demonstrate poleward surges seeing in the remnant magnetic flux and associated bright Ca~II~K plages \citep{Sheeley11, Bertello20}.  They also found that intense remnant flux surges resulted in the polar field reversals. 
As ARs evolve and decay, plages are observed in the
chromosphere. Chromospheric filaments that trace the polarity
inversion lines also outline large-scale magnetic patterns. Long-term
H$\alpha$
observations were compiled in the synoptic maps of
large-scale magnetic fields \citep{McIntosh1979}.  \citet{Makarov1983} analyzed similar
synoptic maps for 1945-1981. Using H$\alpha$ synoptic maps from
Kodaikanal Solar Observatory (KoSO) they studied filament poleward
migration for Cycles 15--21. Their approach, however, does not
provide any information on magnetic field strength. A good correlation
between the Ca~II~K line intensity of chromospheric plages and the
unsigned magnetic flux density \citep{Skumanich75} was used to create pseudo-magnetograms
of ARs \citep{Pevtsov2016}. Reconstruction methods were also developed using historical
data on the polarities of magnetic fields in sunspots \citep{Pevtsov2016}. This
reconstruction reproduces large-scale magnetic patterns, including
the poleward flux surges and the strength of polar fields. The
shortcoming of this approach is that magnetic polarities of the
weak ARs remain undefined. It is also difficult to identify remnant
fluxes related to decay of complex or anomalous ARs.

We develop a new method of the Sun's magnetic field reconstruction
by treating the Ca~II~K intensity and H$\alpha$ synoptic maps from
KoSO \citep{Priy14,Chatterjee2016} as characteristics of the magnetic flux strength and its
polarity. 
The signs of the reconstructed magnetic flux are assigned according to
the polarities derived from the corresponding H$\alpha$ synoptic
maps. Due to no changes in the instrumentations over a century in
the KoSO, the uniform data quality allowed us to reconstruct a
homogeneous magnetic field data of the Sun for five complete solar
cycles (Cycles 15--19) for the first time when no direct magnetic
field observations are available.

\section{Model and Data analyses}
\label{sec:mod}
\subsection{Magnetic flux and its proxies}
\begin{figure}
\centering
\includegraphics[scale=0.42]{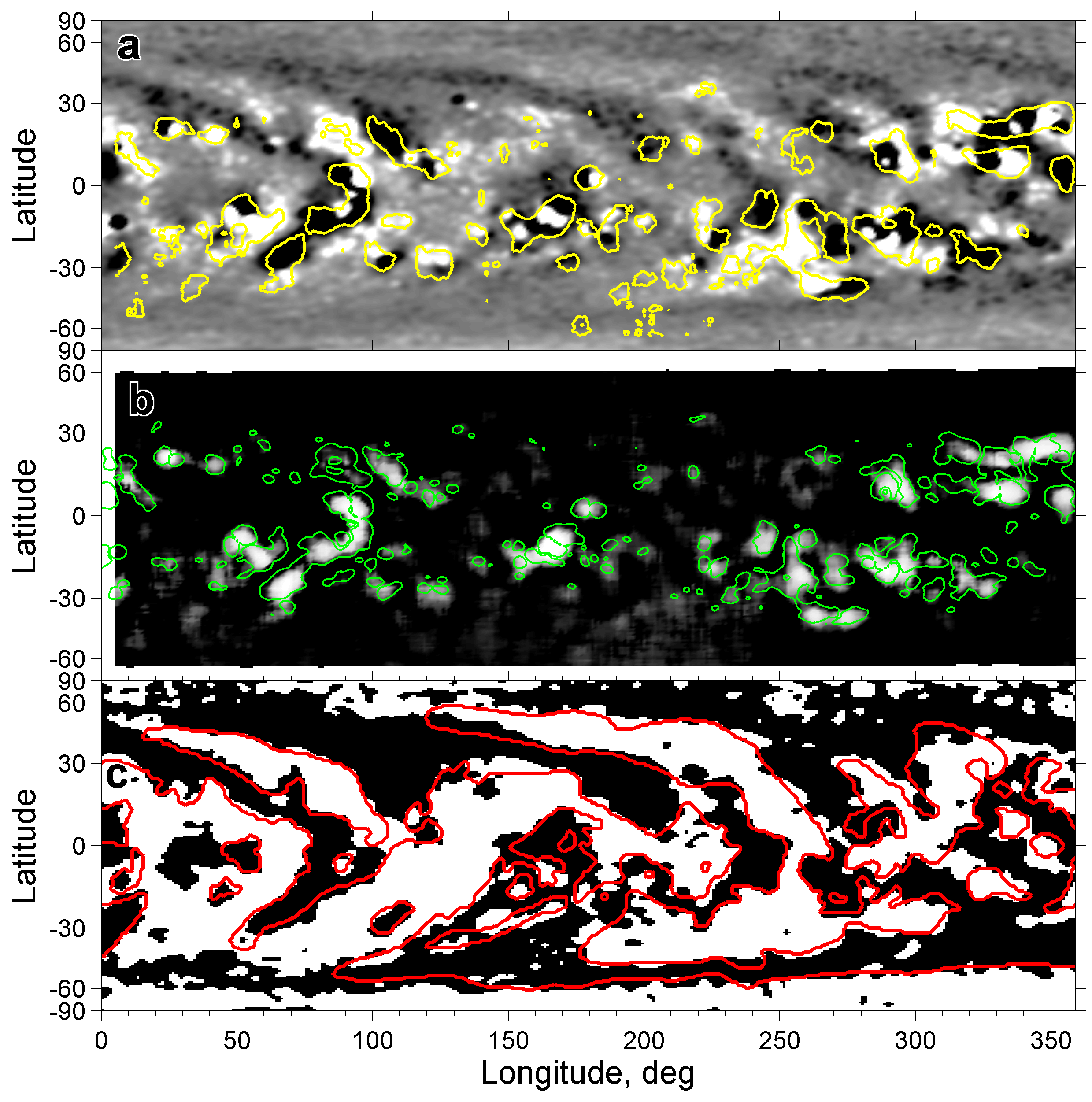}
\caption{
Synoptic map for CR 1717.  (a) Magnetic flux density with positive and negative polarities are shown by black and white, respectively. The yellow contours depict the locations of Ca~II~K plages that exceed a threshold of 170. 
(b) The Ca~II~K intensity map. The green contours show domains of magnetic flux density that exceed 20~G in modulus. 
(c) Distribution of dominant magnetic polarities shaded in black and white. Red contours show the polarity inversion lines derived from the H$\alpha$ map.  
}
\label{fig:synmap}
\end{figure}

Ca~II~K line intensity provides a good proxy for magnetic flux
density 
\citep{Skumanich1975,Mandal2017, Chatzistergos19}.
We use daily synoptic
maps of Ca~II~K and H$\alpha$ that were recently digitized in 16-bit
KoSO native scale \citep{Priyal2014,Chatterjee2016,Chatterjee2017}.
We consider all datasets in the Carrington coordinate system to
examine the related values pixelwise. 
However, due to issues with the
conditions of the photographic plates and a large number of missing days
during the recent cycles, \citep[see Figures 1(b) and 4 of][]{Chatterjee2016}, 
we choose to limit our analysis to the period between 1907 and 1965,
which covers from Cycle~14 (descending phase only) to Cycle~19.
To demonstrate the correlation
with the magnetic flux, we use synoptic magnetograms from Kitt Peak
Observatory \citep{Jones1992}. 
\Fig{fig:synmap}a shows synoptic maps of
magnetic flux for Carrington Rotation (CR) 1717. Positive/negative polarities are shown
in white/black. Yellow contours depict the Ca~II~K plages that exceed
a threshold of 170 in the KoSO native scale. The KoSO Ca~II~K intensity
synoptic map is shown in \Fig{fig:synmap}b. Green contours show patterns of
magnetic flux density that exceed 20~G in modulus.

\citet{McIntosh1979} constructed H$\alpha$ synoptic maps using daily
spectroheliograms. H$\alpha$ filaments and filament channels represent
polarity inversion lines that trace boundaries of the large-scale
patterns above underlying photospheric UMRs. The H$\alpha$ synoptic
maps display polarity inversion lines that outline boundaries of
large-scale magnetic field \citep{Makarov1983,Pevtsov2016,Webb2018}. 
The H$\alpha$ maps agree with
synoptic magnetograms \citep{Snodgrass2000}. \Fig{fig:synmap}c shows dominant magnetic polarities
in black and white. This distribution was estimated using the
majority function of morphological mathematics. The most extended
surges approach the Sun's poles. Red contours show polarity inversion
lines derived by \citet{McIntosh1979} for the same CR. It should be noted
that the distribution obtained from the H$\alpha$ synoptic map is
in agreement with the results of direct measurements (\Fig{fig:synmap}a).

The H$\alpha$ synoptic maps for Cycle 19 were presented in \citet{Makarov1986}, 
followed by a complete data set for Cycles 15--19 \citep{Makarov2007a,Makarov2007b}.
We digitized the H$\alpha$ synoptic maps with one-degree step both
in longitude and latitude. These maps cover CRs 815--1486. For ease
of comparison, we represented the H$\alpha$ digital maps uniformly
over the sine latitude. The H$\alpha$ synoptic maps quantify
distributions of dominant magnetic polarities with values $\pm1$ according
to the usual convention of polarity signs \citep{McIntosh1979}. Thus, synoptic maps
of the Ca~II~K line intensity correlate well with the unsigned
magnetic flux. We also use H$\alpha$ synoptic maps in the form that
indicates distributions of dominant magnetic polarities which are
coded by $\pm1$. All synoptic maps are of the same size $180\times360$ with
uniform sampling over the longitude and sine of latitude.

\subsection{Empirical reconstruction of magnetic flux}
We study the regression relation between the Ca~II~K intensity and unsigned magnetic flux density over the period of 1975--1992 when both datasets are available. To estimate the intensity-flux relation, we selected synoptic maps for CRs 1638, 1694, 1717, 1731, and 1754. They are characterized by good quality and lack of data gaps, and importantly these synoptic maps cover different phases of the solar cycle.
The regression relation between the Ca~II~K intensity and magnetic flux density is non-linear and looks very noisy \citep{Pevtsov2016}. To reduce a scatter of the regression analysis, we smoothed every synoptic map using a median filter over a mask of $11\times7$ pixels. Such a robust smoothing of both Ca~II~K intensity and magnetic flux is appropriate to represent a typical AR and its chromospheric effect. We believe that this approach is suitable to identify plages, their large-scale structure, and the magnetic network contribution. However, such a reconstruction is unable to reproduce subtle details in the distribution of the Ca~II~K intensity.

\begin{figure}
\centering
\includegraphics[scale=0.11]{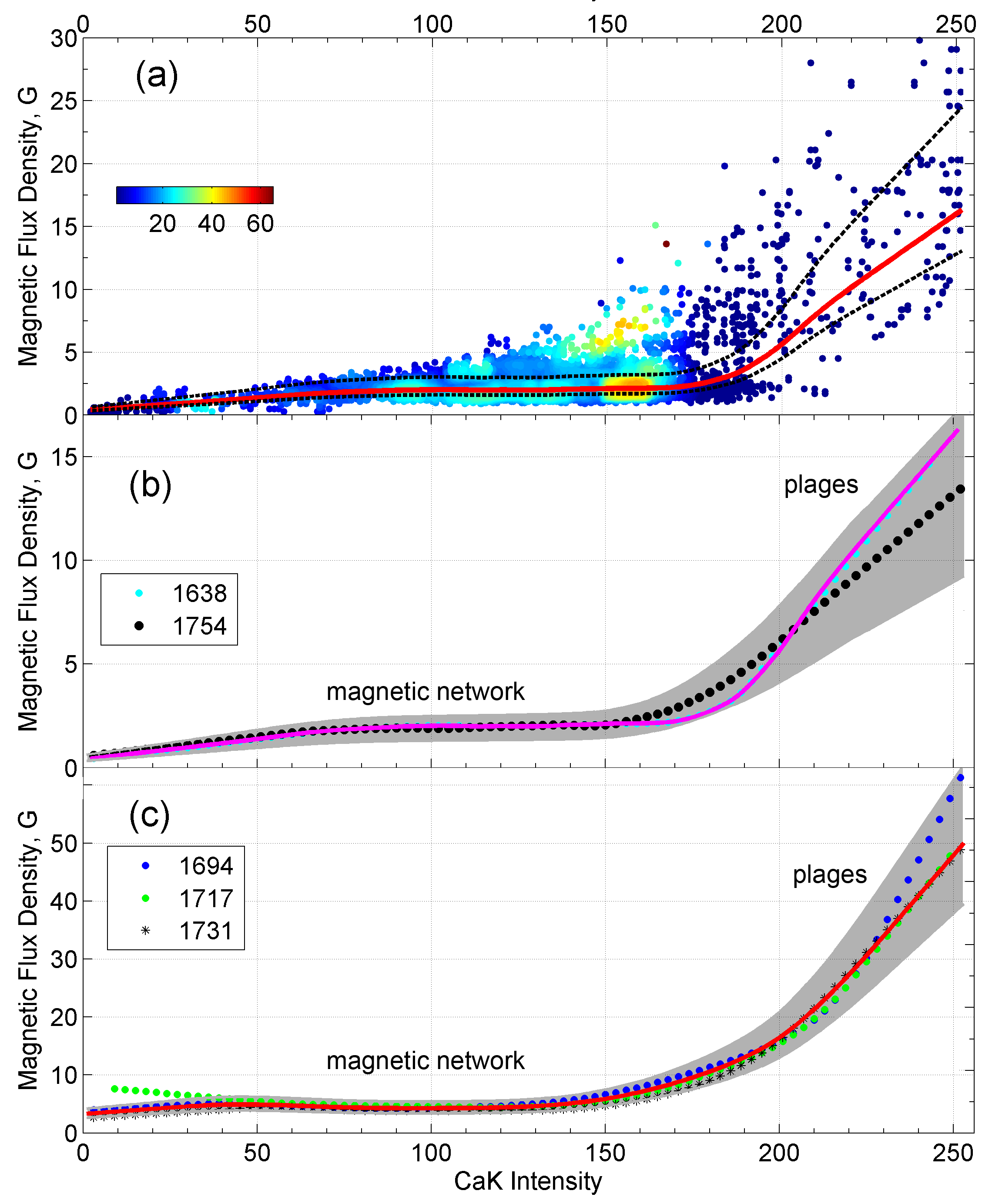}
\caption{
The magnetic flux density dependence on the Ca~II~K the intensity. (a) For CR 1638 in which
color markers and spots represent data density. The empirical regression is shown 
by the red curve and black dotted ones show its upper and lower limits.
(b) For CRs~1638 and 1754 (activity minima) are shown in cyan and black. 
The generalized regression at the activity minima is shown in magenta. 
(c) Similar dependencies for CRs~1694, 1717, and 1731 (around activity maximum of Cycle 21)
are shown in in blue, green, and black, respectively, while the generalized regression are shown in red.
The error corridors are shown in gray. 
}
\label{fig:intvsflux}
\end{figure}

When calculating the regression relation, it is important to use robust algorithms that are resistant to large errors in data. We use the Locally Weighted Scatterplot Smoothing (LOWESS) algorithm \citep{Cleveland1979} to estimate non-linear regression relation between smoothed synoptic maps pixel-to-pixel. Solar activity was low during 1976.11--1976.18 and 1984.76--1984.84, which correspond to CRs 1638 and 1754, while CRs 1694 (1980.28--1980.36), 1717 (1982.00--1982.08), and 1731 (1983.05--1983.12) occurred in high activity phase (around solar Cycle 21 maximum).
\Fig{fig:intvsflux}a shows the scatter plot that represents the intensity-flux density dependence for CR 1638.  
The markers and dots show data density in the intensity-flux density space.
Despite its scattered distribution, the LOWESS algorithm makes it possible to evaluate a robust empirical regression at the $95\%$ confidence level (red curve). The upper and lower limits (black dotted curves) depict the corridors of possible errors in the regression estimate. These corridors also characterize the errors in magnetic flux reconstruction. The errors are about $\pm2$ G at low Ca~II~K intensities ($<150$). The reconstruction errors increase rapidly with increasing intensity, reaching about $\pm 10$ G at the highest intensity.

Using the LOWESS algorithm, we estimated the regression relations for the selected CRs. They are shown in \Fig{fig:intvsflux}b-c.  All curves are grouped into two types, which depend on magnetic activity levels. The regression curves for CRs 1638 and 1754 are shown in cyan and black in \Fig{fig:intvsflux}b. 
By combining these curves as subsets of data points, we can define a generalized regression model for the intensity-flux relation during activity minima. Using the LOWESS algorithm, we determine a generalized regression relation that is shown in magenta. A similar analysis is also applied for CRs~1694, 1717, and 1731 that occurred at the high magnetic activity (around Cycle 21 maximum) which are shown in \Fig{fig:intvsflux}c.

It should be noted that the regression relation is composed of two parts. For intensities of less than 180, magnetic flux increases slowly, while it increases rapidly at higher intensities. The low-slope dependence represents a quiet chromosphere and magnetic network. As an 11-year cycle progresses, the contribution of the magnetic network varies. The contribution of the strong magnetic field appears with higher intensities and the dependence becomes steep. Bright plages usually appear above ARs and are associated with their remnant flux.
We can approximate these generalized regression relations using adequate polynomial models. 
The polynomial models of the third (\Eq{eq1}) and fourth orders (\Eq{eq2}) fit the regression relations at low- and high-activity levels. Their regression equations are
\begin{eqnarray}
F = 2.7801\times10^{-6} C^3 - 0.00070527 C^2 + 0.059549 C \nonumber  \\
+ 0.022785,
\label{eq1}
\end{eqnarray}

\begin{eqnarray}
F = 2.0656\times10^{-8} C^4 - 7.0741\times10^{-8}C^3  \nonumber  \\
 - 8.3844\times10^{-4} C^2 + 0.07576 C + 3.0329.
\label{eq2}
\end{eqnarray}
Here unsigned magnetic flux density ($F$) and the correspondent Ca~II~K intensity ($C$) are taken pixel-wise from synoptic maps. The error corridors are shown in grey for every model in \Fig{fig:intvsflux}b-c.
Taking into account magnetic activity levels, we select an appropriate model for the reconstruction of unsigned magnetic flux. 
Applying the regression models (\Eqs{eq1}{eq1}) to the Ca~II~K synoptic map (pixel by pixel), we reconstruct the unsigned magnetic flux density for CRs 815--1486. The H$\alpha$ synoptic maps from KoSO determine the polarity of the magnetic flux density. 
We note that the measurements of magnetic fields sometimes differ significantly with that inferred from H$\alpha$ synoptic maps, 
especially in the polar regions \citep{Snodgrass2000}. However, such errors are comparable to those from regression relations. 
The main source of the error in our reconstructed magnetic field is in constructing the regression relation between the Ca~II~K intensity and the magnetic flux.  The error in our reconstructed field is $\pm1$ G at the weak field and reaches to $\pm10$ G at the strongest field. 

\begin{figure*}
\centering
\includegraphics[scale=0.12]{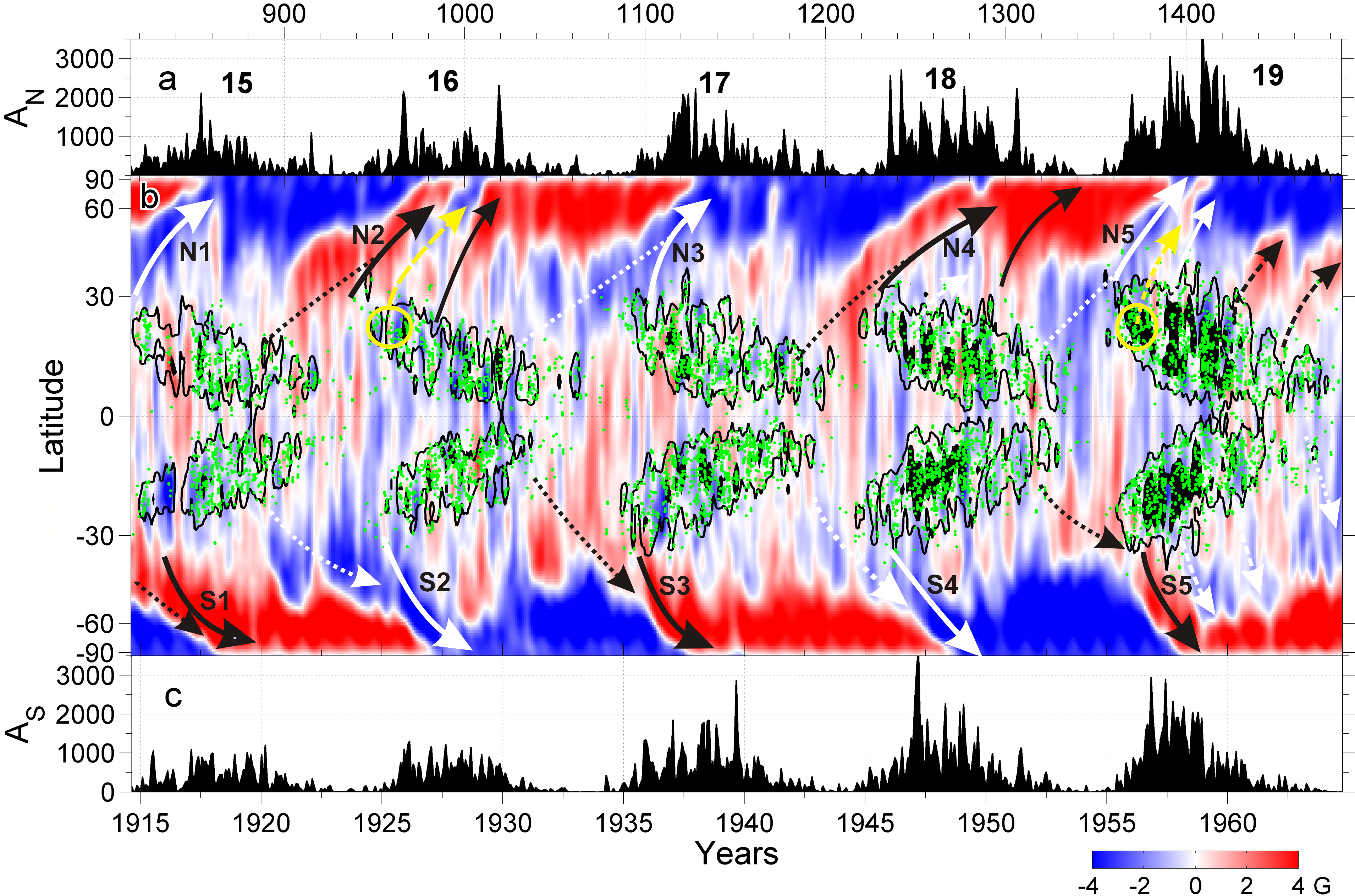}
\caption{
Butterfly diagram of reconstructed magnetic field on the solar surface. (a),(c) Temporal variation of the sunspot areas in the northern/southern hemisphere. (b) Zonally averaged magnetic field is shown in blue-to-red. Zones of intense sunspot activity, domains of negative tilt predominance are shown in black and green, respectively. Numbers of solar cycles are labeled at the top panel. The remnant flux surges of following/leading polarities are shown by solid/dashed arrows. Critical surges are shown with bold arrows and labeled as N1/S1, N2/S2, N3/S3, N4/S4, and N5/S5 for Cycles 15--19. The yellow ovals show possible sources of the (opposite) leading-polarity surges.
}
\label{fig:bfly}
\end{figure*}

\section{Results and Discussion}
\label{sec:res}
To study long-term evolution of the Sun's magnetic field in detail 
 we applied the time-latitude analysis to the synoptic maps of magnetic flux. Successive maps were zonally averaged for every CR. The time-latitude distribution of magnetic flux was denoised using the non-decimated wavelet decomposition. \Fig{fig:bfly}b depicts the zonally averaged magnetic field in blue-to-red shading.
The time-latitude distribution, the so-called butterfly diagram of the zonal-averaged reconstructed magnetic field is shown in \Fig{fig:bfly}b. To show the variation of the solar cycle strength, the sunspot area time series (which gives a rough estimate of the total magnetic flux on the solar surface) in each hemisphere is shown in \Fig{fig:bfly}a and c. Basic features of the solar radial magnetic field evolution on the surface, namely, the polarity reversal, the predominant dipolar nature near solar minima, and a poleward migration of the dominant magnetic polarities from mid- to high-latitudes, are clearly observed. The details of the polar field development are also observed in \Fig{fig:bfly}b. As the ARs evolve, their fluxes decay within the band of intense sunspot activity as shown by black contours.

The decay of these ARs results in remnant flux surges, which are transported polewards. The major remnant flux surges of the following/leading polarities are shown with solid/dashed arrows. As the extended surges approach the Sun's polar regions, reconnection of the ‘old’ magnetic flux with the `new' one starts. The latitudinal extent of the old flux decreases. The polar field reversal occurs at complete cancellation of the opposite polarity field. The further transport of the following polarities UMRs leads to buildup of the ‘new’ polar fields. This polar field reaches its maximum strength around the solar minimum. It takes about 1.5--2 years for the remnant field to reach the Sun's poles. The poleward surges of remnant flux result in the polar field reversals. The first surges that reached the Sun's poles are of crucial importance because they led to the polar-field reversals.  In \Fig{fig:bfly}b, the critical surges are shown with bold arrows. In the northern/southern hemispheres, they are labeled as N1/S1, N2/S2, N3/S3, N4/S4, and N5/S5 for Cycles 15--19, respectively. In both hemispheres, the polar-field reversals sometimes occur asynchronously. \Tab{table1} depicts the polar-field reversal timing in both hemispheres for Cycles 15--19. 
On comparing the results from \citet{MS86}, we find that our values of the timings of the polarity reversals 
are very close to their values. 
Further, in agreement with \citet{MS86}
Cycles 16 and 19 had triple polarity reversals for the northern hemisphere, which was also seen in Cycles 20 and 21 \citep{Mordvinov2019}.

The SFT \citep{Jiang2014,Bhowmik2018} as well as the Babcock-Leighton dynamo \citep{KM17,KM18,LC17} models have shown that the scatter in the BMR tilt around Joy’s law produces changes in the polar field. Particularly, the anti-Hale and non-Joy BMRs cause remnant flux surges of the opposite (leading) polarities that disturb the usual (following) polarities of the magnetic field. To identify the causal connection between the tilt of BMRs and the polar field, we have studied the sunspot group tilts measured at KoSO \citep{Sivaraman1999}. We selected ARs with negative tilts from the KoSO archive and indicated their positions with green markers in \Fig{fig:bfly}b. In each cycle, there are domains of negative tilts. We selected two such domains that are shown by yellow ovals in the northern hemisphere in Cycles 16 and 19. These domains also cover bands of intense sunspot activity. We observe that the opposite (leading) polarity surges are triggered from these bands of negative tilt (shown by dashed/yellow arrows). In Cycles 16 and 19, the leading-polarity surges reached the North Pole and caused the second reversals.

It is usually believed that the total magnetic flux that emerges during an 11-year cycle reconnects at the Sun's polar zones and near the equator by the end of each cycle. The time-latitude analysis revealed also unusual patterns that link adjacent cycles. By the end of every cycle, UMRs of the leading polarities are usually formed at low- and mid-latitudes. During activity minima, these UMRs are transported to higher latitudes due to the meridional flow. As the next cycle progresses, these UMRs merge with following polarity surges of the new cycle which are transported to the Sun's polar zones. Such patterns are well-defined in Cycles 15--19. They are shown with dotted arrows in \Fig{fig:bfly}b. These surges show new links between the adjacent solar cycles. It is the interrelation that combines individual 11-year cycles as parts of the Hale (22-year) cycle. The merger of the UMRs of the new and previous cycles leads to an increase of the poleward flux transport that results in a strengthening of polar magnetic field in the subsequent cycle. This interrelation between 11-year cycles suggests a possible long-term memory in solar activity that may vary on a secular timescale.

\begin{table}
\caption{
Timings of the polar-field reversals in the north and south poles for Cycles 15--19.
The corresponding values from \citet{MS86} (MS86) are also shown for the comparison.
}
\begin{center}
\begin{tabular}{lcccc}
\hline
\hline
Cycle & \multicolumn{2}{c}{North}  & \multicolumn{2}{c}{South} \\
\cline{2-3}
\cline{4-5}
number& Our value   & MS86   & Our value   & MS86                \\
\hline
15    & 1917.4 & 1918.6  & 1918.2 & 1918.7\\
16    & 1927.2,& 1927.9, & 1927.0 & 1928.5\\
      & 1928.6,& 1929.3, &        &       \\
      & 1929.6 & 1929.9  &        &       \\
17    & 1938.0 & 1940.1  & 1938.1 & 1940.0\\
18    & 1950.3 & 1950.2  & 1948.6 & 1949.0\\
19    & 1957.5,& 1958.0, & 1958.2 & 1959.5\\
      & 1958.8,& 1958.8, &        &       \\
      & 1959.4 & 1959.7  &        &       \\
\hline
\end{tabular}
\end{center}
\label{table1}
\end{table}

%

\begin{figure}
\centering
\includegraphics[scale=0.34]{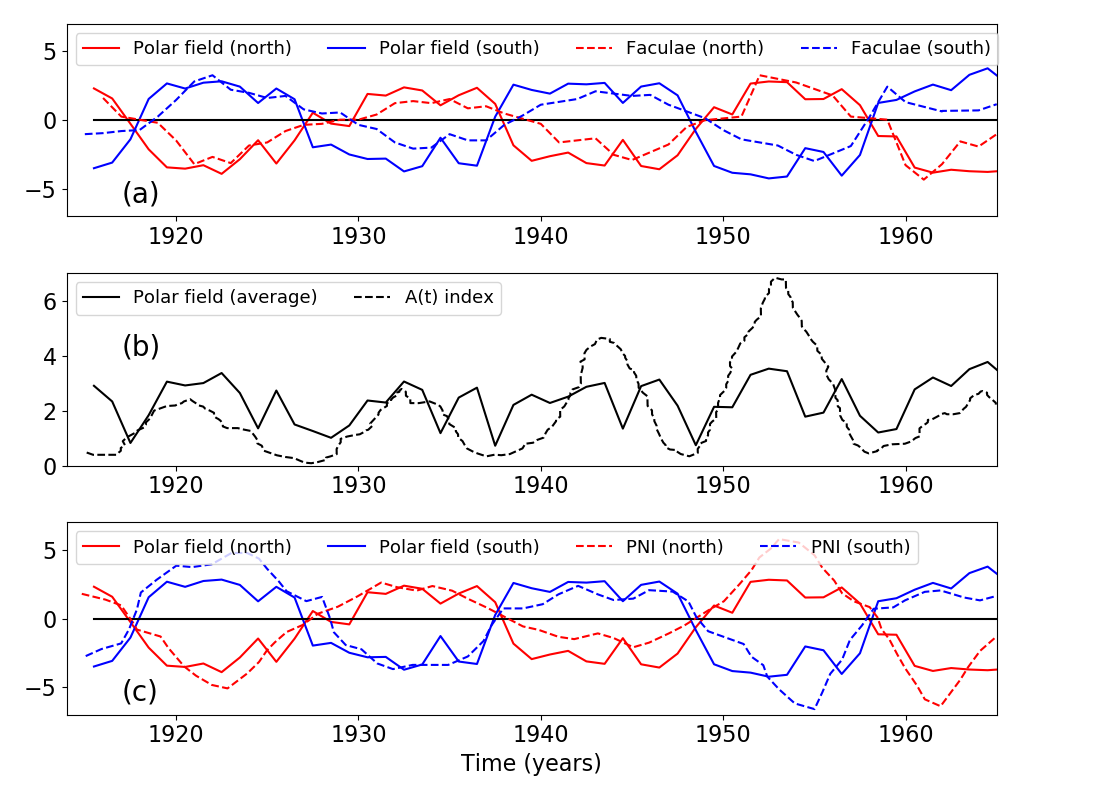}
\caption{
Comparison of our reconstructed polar magnetic field and the polar field proxies. 
(a), (c) Solid lines: Temporal variations of the polar field computed by averaging 
the reconstructed magnetic field from $55^\circ$ to pole (red/blue: north/south). 
(a) Dashed lines: The polar field proxy as inferred from the polar faculae count. 
Note the Spearman rank correlations between our reconstructed polar field and the polar faculae data
are $0.88 (p = 0.0)$ and $0.74 (p = 0.0)$ for the northern and southern hemispheres, respectively.
(b) Solid line: the average of north and south absolute polar field, dashed line: Makarov's $A(t)$ index. 
(c) Polar Network Index (PNI) (red/blue: north and south).
}
\label{fig:comp}
\end{figure}

Finally, to assess a quality of the magnetic field reconstruction in a global aspect, we estimated the average values of the magnetic flux density in the Sun's polar regions and compared them with indirect data obtained independently.
We find a statistically significant correlation between our reconstructed polar field and the other available proxies of the polar field, such as polar faculae \citep{Munoz2012}, $A(t)$ index \citep{Makarov2001} and Polar Network Index (PNI) \citep{Priyal2014}; see \Fig{fig:comp}.
A good correlation of polar magnetic fields and indirect indices of activity confirms the satisfactory quality of the reconstruction.

\section{Summary and Conclusions}
In summary, using the homogeneous proxies of the magnetic field from KoSO, we have reconstructed the magnetic field on the solar surface for Cycles 15--19. Our results enable us to identify and explore the cause of the polar field development and the polarity reversals, including the multiple reversals in some cycles. These findings improve our understanding of the generation of the solar magnetic field and the cycle. We believe that our result of the reconstructed magnetic field will be utilized to constrain the parameters of the SFT and dynamo models.

\begin{acknowledgements}
We thank the anonymous referee for careful reading and offering valuable comments, suggestions and referenes, all these helped to improve the quality of the paper.
This work was supported by Russian Foundation for Basic Research project 19-52-45002$\_$Ind under the Indo-Russian 
Joint Research cooperation (from Russian side) and Indo-Russian Joint Research Program of Department of Science and Technology with project number INT/RUS/RFBR/383 (Indian side).
The work was performed with budgetary funding
of Basic Research program II.16.
Financial supports from Department of Science and Technology
(SERB/DST), India through the Ramanujan Fellowship (project no
SB/S2/RJN-017/2018) and ISRO/RESPOND (project no ISRO/RES/2/430/19-20)
are also acknowledged. 
\end{acknowledgements}

\bibliographystyle{apj}
\bibliography{paper}

\end{document}